\documentclass[english,aps,prd,superscriptaddress,preprintnumbers,nofootinbib,preprint]{revtex4}
\usepackage[T1]{fontenc}
\usepackage[latin9]{inputenc}
\usepackage{verbatim}
\usepackage{amsmath}
\usepackage{dcolumn}
\usepackage{graphicx}
\newcommand{\be}{\begin{equation}}
\newcommand{\ee}{\end{equation}}
\newcommand{\bea}{\begin{eqnarray}}
\newcommand{\eea}{\end{eqnarray}}

\newcommand{\mrm}{\mathrm}

\def\sigmabkgdLO{\sigma^{\mathrm{bkgd}}_\mathrm{LO}}
\def\sigmafullLO{\sigma^{\mathrm{full}}_\mathrm{LO}}
\def\sigmasiglLO{\sigma^{\mathrm{signal}}_\mathrm{LO}}
\def\sigmaintfLO{\sigma^{\mathrm{intf}}_\mathrm{LO}}
\def\sigmabkgdNLO{\sigma^{\mathrm{bkgd}}_\mathrm{NLO}}
\def\sigmafullNLO{\sigma^{\mathrm{full}}_\mathrm{NLO}}
\def\sigmasiglNLO{\sigma^{\mathrm{signal}}_\mathrm{NLO}}
\def\sigmaintfNLO{\sigma^{\mathrm{intf}}_\mathrm{NLO}}
\def\m4l{m_{4\ell}}
\def\myell{\ell}

\usepackage{babel}
\usepackage{color}
\begin{document}
\global\long\def\order#1{\mathcal{O}\left(#1\right)}
\global\long\def\d{\mathrm{d}}
\global\long\def\P{P}
\global\long\def\amp{{\mathcal M}}
\preprint{TTP-16-017, CERN-TH-2016-119}

\def\KIT{Institute for Theoretical Particle Physics, KIT, Karlsruhe, Germany}
\def\CERN{CERN Theory Division, CH-1211, Geneva 23, Switzerland}

\title{QCD corrections to vector boson pair production in gluon fusion including 
interference effects with off-shell Higgs at the LHC}

\author{Fabrizio Caola}            
\email[Electronic address: ]{fabrizio.caola@cern.ch}
\affiliation{\CERN}

\author{Matthew Dowling}            
\email[Electronic address: ]{matthew.dowling@kit.edu}
\affiliation{\KIT}

\author{Kirill Melnikov}            
\email[Electronic address: ]{kirill.melnikov@kit.edu}
\affiliation{\KIT}

\author{Raoul R\"ontsch}            
\email[Electronic address: ]{raoul.roentsch@kit.edu}
\affiliation{\KIT}

\author{Lorenzo Tancredi}            
\email[Electronic address: ]{lorenzo.tancredi@kit.edu}
\affiliation{\KIT}

\begin{abstract}
We compute next-to-leading order (NLO) QCD corrections  to the 
production of two massive electroweak bosons  in  gluon fusion. 
We consider both the prompt production process $gg \to VV$ and 
 the production mediated by an exchange of an $s$-channel  Higgs boson,  $gg \to H^* \to VV$. 
We include final states with both on- and 
off-shell vector bosons with leptonic decays.
The gluonic production of vector bosons is a loop-induced process, 
including both massless and massive quarks in the loop.
For $gg \to ZZ$ production, we  obtain the NLO QCD corrections to the massive loops through an expansion 
around the heavy top limit. 
This approximation is valid below the top production threshold, 
giving a broad  range  of invariant masses  between the Higgs production and the top production 
thresholds in which our results are  valid.  We explore the  NLO QCD 
effects in $gg \to ZZ$ focusing, in particular, on the  
interference  between  prompt and Higgs-mediated  processes. We find 
that the QCD corrections to the interference are large and similar in  size to the corrections 
to both the signal  and the background processes.   
At the same time, we observe that  corrections to the interference change 
rapidly with the four-lepton invariant mass in the region around 
the $ZZ$ production threshold.  We also study the interference effects in $gg \to W^+W^-$ production 
where, due to technical limitations, we  only consider contributions 
of massless loops. 
We find that the QCD corrections to the interference in this case 
are somewhat 
larger than those for either the signal or the background.
\end{abstract}

\maketitle

\section{Introduction} 
After the discovery of the Higgs boson during Run I 
at the Large Hadron Collider (LHC)~\cite{Aad:2012tfa,Chatrchyan:2012xdj} an important 
task for Run II is a thorough study of its properties.
In the Standard Model (SM), the Higgs field  is solely 
responsible for the phenomenon of electroweak symmetry breaking (EWSB) that  
provides   masses to  fermions and weak gauge bosons in a consistent way. 
This minimal version of the EWSB  mechanism predicts a stringent relation between masses 
of elementary particles and their couplings to the Higgs boson.
Experimental studies 
of Higgs couplings to other Standard Model particles  
provide a direct test of this  mechanism; any deviation from the minimal relation 
between couplings and masses will imply that the SM version of the EWSB mechanism is 
incomplete. 

Measurements of the couplings are typically performed for the on-shell production and decay 
of the Higgs bosons, simply because  the absolute majority of 
the Higgs bosons at the LHC  are produced on-shell. 
However, it was recently   realized that the off-shell production  
of the Higgs boson can also provide useful insights into its properties. Indeed, 
despite the extremely narrow width of the Higgs boson, the off-shell 
region is well-populated, with about one out of ten  Higgs boson events in the $gg \to H \to ZZ$ 
process   having an invariant mass of the two $Z$-bosons 
above  $180~{\rm GeV}$~\cite{Kauer:2012hd}.
Furthermore, the interference between Higgs-mediated amplitude 
$gg \to H \to VV$ and the prompt production amplitude  $gg \to VV$  
is strong and destructive in the high invariant mass region. 
This is expected since Higgs boson exchanges are supposed to unitarize the scattering 
amplitudes of massive fermions and gauge bosons.  The observation of this unitarization effect at large values 
of $m_{ZZ}$ will be an important confirmation of the fact that the discovered Higgs boson 
is indeed the only  agent of electroweak symmetry breaking mechanism,  as predicted in the Standard Model. 

Measurements in the off-shell region are also useful for another reason. Indeed, extraction of the 
Higgs couplings from the on-shell measurements is, in principle, compromised by the unknown 
value of the Higgs width, leading to an infinite-fold degeneracy in the extracted values of the 
couplings.  It was pointed out by two of us \cite{Caola:2013yja} that this ambiguity is lifted by off-shell 
measurements, which are sensitive to the Higgs couplings only. The 
ratio of the off-shell to on-shell cross sections  $ pp \to H \to ZZ$ 
can then be used to obtain  stringent constraints on the Higgs boson width~\cite{Caola:2013yja,Campbell:2013una,Campbell:2013wga}.
Subsequent experimental analyses used  this method  to constrain  the Higgs boson width 
to be less than a few times its Standard Model value~\cite{Khachatryan:2014iha,
Aad:2015xua,Khachatryan:2015mma}.
This is to be compared with the {\it direct} constraints on the Higgs width that 
can be obtained from fitting the invariant mass distributions of four leptons or a photon pair around the Higgs 
mass. Because of detector resolution, such direct constraints on $\Gamma_H$ cannot probe values smaller
than $\Gamma_H\sim 1~{\rm GeV}$, so they are about two orders 
of magnitude weaker than the indirect constraints based on off-shell measurements~\cite{CMS:2013wda}.

The constraints on the Higgs width obtained from the off-shell measurements are 
not model-independent; the primary assumption is that  the effective 
Higgs couplings to SM particles  do not  differ substantially in the 
on-shell and off-shell regions~\cite{Englert:2014aca,Englert:2014ffa,Logan:2014ppa}. 
 There are several  ways to 
make this assumption invalid. For example, one can  extend  the theory to include 
relatively light colored particles that  contribute to the  $ggH$ coupling~\cite{Englert:2014ffa},
new Higgs resonances~\cite{Logan:2014ppa} or anomalous $HZZ$ couplings~\cite{Anderson:2013afp,Gainer:2013rxa,Gainer:2014hha}.
However,  all such  cases will give rise to relatively clear signatures 
of New Physics at the LHC,  beyond  changes in the Higgs 
width or a change in the number of events in the off-shell region. 
As the result, the validity of the assumptions crucial for 
 the extraction of the Higgs width from the comparison of off- and on-shell Higgs 
production cross sections can be experimentally validated. 

However, if  the 
couplings of the Higgs boson to gluons and vector bosons are modified,  the yield of vector bosons $V$, produced 
in gluon fusion,  changes in a complicated  way. 
This is because the Higgs signal $gg \to H \to VV$ amplitudes
and their interference with the prompt production amplitudes scale differently. 
As a result, it is important to investigate properties of the signal $gg \to H \to VV$, 
the irreducible background $gg \to VV$, and the interference separately.  In particular, the QCD corrections 
to {\it each} of these contributions should be known  since this information is required  to properly 
simulate the $gg \to ZZ$ process with modified Higgs couplings.

A significant amount of recent effort has been 
focused on QCD corrections to both Higgs and 
massive $VV$ production, 
resulting in the former being computed to 
next-to-next-to-next-to leading order (N$^3$LO) in QCD in the heavy 
top limit~\cite{Anastasiou:2015ema,Anastasiou:2016cez} and the latter 
to next-to-next-to leading order (NNLO)~\cite{Cascioli:2014yka,Gehrmann:2014fva,Grazzini:2015hta,Grazzini:2016swo,Grazzini:2016ctr}.
By contrast,  although the gluonic prompt background (with on-shell $Z$) is known through next-to-leading order (NLO) in QCD~\cite{Caola:2015psa,Caola:2015rqy},
the interference has so far only been computed at leading order (LO). This is unfortunate since the QCD corrections 
to gluon-induced processes are known to be significant,   partially  due to the high likelihood of emitting 
a  hard gluon in the processes $gg \to H$ and $gg \to VV$. 
In the current experimental analyses,  the QCD enhancement of  the interference effects is modeled approximately, 
by  assuming that it is related to  the known QCD enhancement of the signal. While this is a plausible 
hypothesis  which can be  justified if universal  soft QCD radiation provides a 
dominant source of radiative corrections~\cite{Bonvini:2013jha,Li:2015jva},  it is important to verify it by an explicit 
computation. Such verification  as well as the 
computation of the realistic QCD enhancement factor for  the interference accounting   
for off-shell effects, vector boson decays and  fiducial cuts  used by ATLAS and CMS
collaborations~\cite{Khachatryan:2014iha,
Aad:2015xua,Khachatryan:2015mma}, 
become  increasingly important since the LHC experiments are posed to push  the 
off-shell measurements to a new level of precision.

Since both signal and background  processes are loop-induced,  QCD computations 
for each of them  require two-loop amplitudes. In the case of 
Higgs production in gluon fusion, the two-loop virtual amplitudes are known since long ago~\cite{Spira:1995rr,Harlander:2005rq,Aglietti:2006tp}.
For the background process $gg \to VV$ the situation is more complex.  Indeed, for both 
neutral and charged vector bosons, $V = Z/\gamma$ and $V = W^\pm$, both massive and massless quarks 
contribute to the two-loop $gg \to VV$ amplitude.  The massless contributions  
were computed during  the last  year~\cite{Caola:2015ila,vonManteuffel:2015msa},
whereas an explicit computation of massive contributions is currently not feasible.
This is because this calculation involves two-loop four-point functions with internal and 
external massive lines, which, despite the recent success in evaluating these amplitudes for the $gg \to HH$ process~\cite{Borowka:2016ehy}, 
 appear to be beyond current loop techniques.
For $V=Z/\gamma$, the problem 
can be circumvented by performing an expansion in the inverse top 
mass as suggested 
 in Ref.~\cite{Melnikov:2015laa}. For $V=W$, the expansion in $1/m_t$ can be done along similar 
lines but it is more complicated
since massive (top) and massless (bottom) propagators appear in contributing diagrams at the same time. 
For this reason we do not compute contributions of third-generation quarks to $gg \to W^+W^-$
in this paper, leaving 
it for future work.

The expansion  of the scattering amplitude $ gg \to ZZ$ 
in $1/m_t$  is expected to be valid 
if the partonic center-of-mass energy is below the  top quark pair production threshold, $\m4l < 2m_t$.
This leaves a significant window of energies $2 m_Z < \m4l < 2m_t$ where  the interference effects are 
important and the heavy-top expansion is expected to provide 
a reasonably  accurate approximation to the massive  $gg \to ZZ$ amplitudes. Computation of QCD corrections to the 
interference for   the invariant masses of the two $Z$-bosons larger than $2 m_t$ 
remains an interesting problem; it can only be fully addressed 
by studying  the NLO QCD corrections to $gg \to ZZ$ amplitudes with the exact mass dependence.

\begin{figure}
\centering
\includegraphics[scale=0.3]{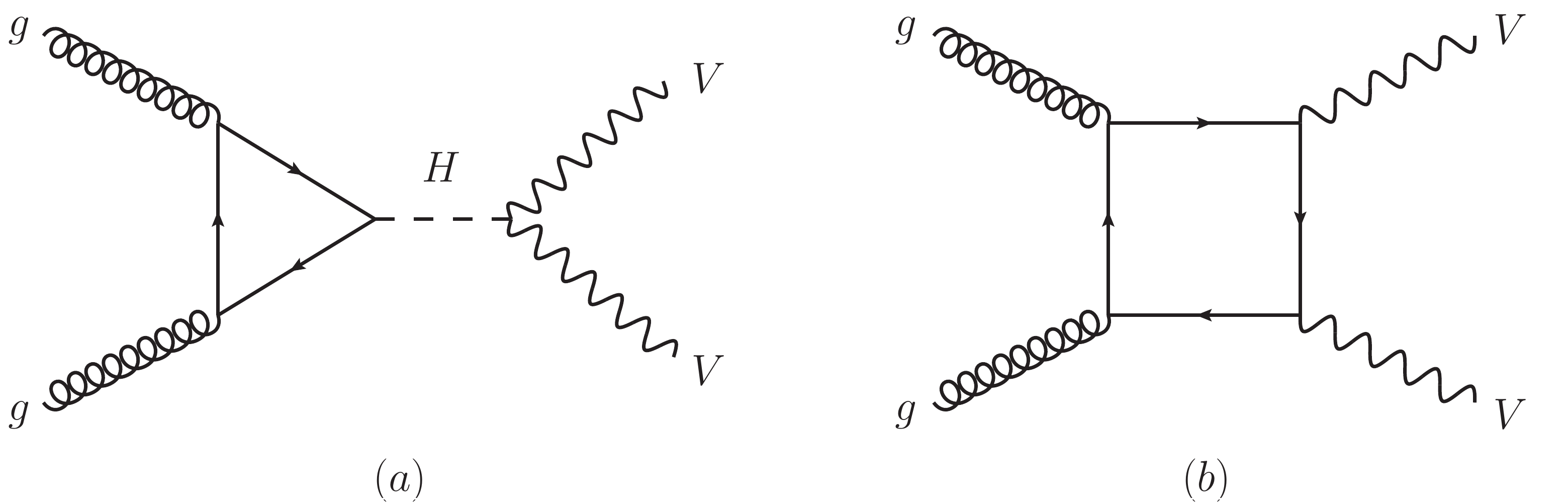}
\caption{Representative Feynman diagrams for the Higgs-mediated signal amplitude $gg \to H \to ZZ$ (a) and the background amplitude $gg \to ZZ$ (b) at LO in pQCD. The decays of the $Z$-bosons to leptons are understood. \label{fig:FeynDiagZZLO}}
\end{figure}

The remainder of this paper is organized as follows. In Section~\ref{sec:ZZ}, we focus on 
$ZZ$ production in gluon fusion. 
We discuss  details of the  calculation, 
including validation of the $1/m_t$ expansion, and present results applicable to the LHC phenomenology.
In Section~\ref{sec:WW}, we present the calculation and discuss 
phenomenology of  the $WW$ production in gluon fusion.
We conclude in Section~\ref{sec:concl}.

\section{$ZZ$ production} \label{sec:ZZ}
\subsection{Details of the calculation} \label{sec:ZZdet}

Scattering amplitudes for processes $ gg \to ZZ$ and $gg \to ZZ + g$ can be written as 
\be
{\cal A}_{\rm ZZ} = {\cal A}_H + {\cal A}_{p},
\label{eq1k}
\ee
where  the first amplitude describes the Higgs-mediated signal process  $gg \to H \to ZZ $ or 
$ gg \to H \to ZZ + g  $   and the second amplitude describes  the ``background'' prompt production 
 $gg \to ZZ $ and $  gg \to ZZ +g $. Although not explicit in these notations, the leptonic decays  
of $Z$-bosons are always included in the calculation and the $Z$-bosons are not assumed to be 
on the mass shell. For background processes, $\gamma^*$-mediated amplitudes are also included.
 Upon squaring the amplitude in Eq.(\ref{eq1k}), one obtains three terms
\be
|{\cal A}_{ZZ}|^2 = |{\cal A}_H|^2 + |{\cal A}_{p}|^2 + 2 {\rm Re} \left [ 
{\cal A}_H^* {\cal A}_{p} \right ],
\label{eq2k}
\ee
that, upon integration  over the phase-space of the relevant final states, produce the corresponding 
contributions to the cross section.  We will refer to the three contributions to the cross sections, 
shown in Eq.(\ref{eq2k}), as the signal, the background and the interference, respectively. Note that 
the interference contribution to the cross section is not sign-definite, in contrast to 
contributions of both the signal and background.

We now describe the ingredients that  we use  to assemble the full scattering amplitude ${\cal A}_{ZZ}$.
The one-loop LO amplitudes ${\cal A}_H$ and ${\cal A}_p$ are shown in Fig.~\ref{fig:FeynDiagZZLO}.
The former,  with full  dependence on the quark masses that facilitate $ggH$ interaction, has been known for a long time. 
The latter amplitudes for both massless and massive quark contributions were computed 
in ~\cite{Glover:1988rg,Matsuura:1991pj,Zecher:1994kb}; more recent computations are available 
in the codes 
$\tt gg2VV$~\cite{Binoth:2008pr} and $\tt MCFM$~\cite{Campbell:2011bn,Campbell:2013una}.
We make use of the amplitudes from $\tt MCFM$ in our calculation. 

\begin{figure}
\centering
\includegraphics[scale=0.4]{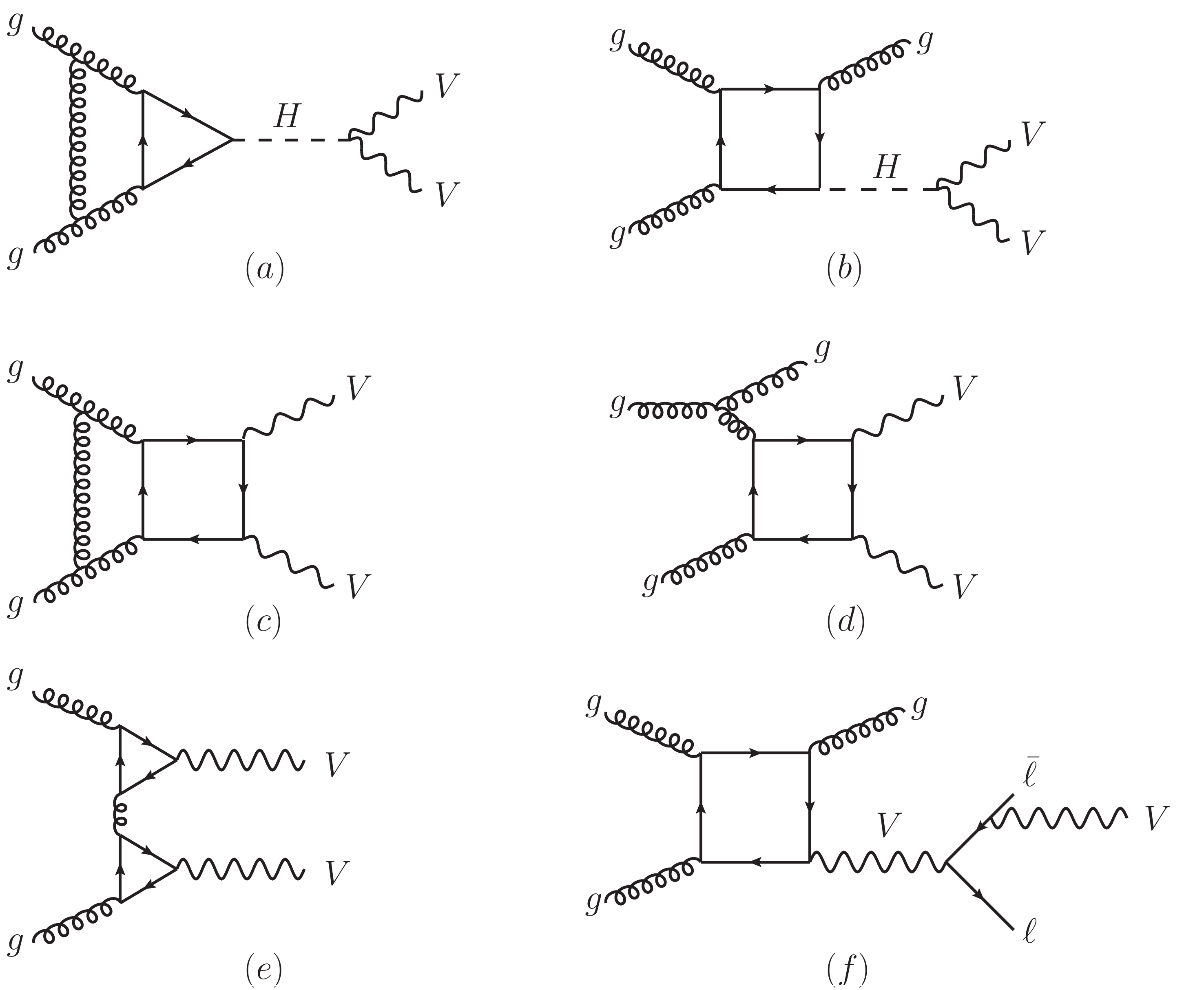}
\caption{Representative Feynman diagrams at NLO. Shown are the two-loop and real emission contributions to the signal amplitude ${\cal A}_{H}$  ((a) and (b)) and to the background amplitude ${\cal A}_{p}$ ((c)-(f)). The decays of the $Z$-bosons to leptons are only shown in (f). \label{fig:FeynDiagZZNLO} }
\end{figure}

For  the NLO QCD computation  we need virtual corrections to $gg \to ZZ$ and real contributions $gg \to ZZ + g$ (see examples of contributing diagrams in 
Fig.~\ref{fig:FeynDiagZZNLO}). To compute the corresponding 
scattering amplitudes, we use the expressions for the two-loop contribution 
to ${\cal A}_H$ from Ref.~\cite{Aglietti:2006tp}. 
The contribution of massless quarks to two-loop background amplitudes ${\cal A}_{p} $ has  been calculated 
in Refs.~\cite{vonManteuffel:2015msa,Caola:2015ila}.
The public libraries of Ref.~\cite{vonManteuffel:2015msa} were already used 
to compute the NLO QCD corrections to the gluon-induced $ZZ$ continuum production~\cite{Caola:2015psa} and 
we borrow the relevant amplitude from that reference.  The NLO QCD corrections 
to the contribution of the top loops to $ZZ$ production 
is not known in an analytic form; we compute it using an expansion in $1/m_t$. The technical details of the 
calculation are  described below.  

The amplitudes for real emission contributions are assembled in a similar way. For the signal process 
$gg \to H + g \to ZZ + g$ the amplitude was computed long ago in Ref.~\cite{Ellis:1987xu,Baur:1988cq}.
For the prompt production  process $gg \to ZZ+g$,  amplitudes that describe the contribution of massless 
quarks 
were calculated using a  combination of analytic and numerical unitarity methods   
in Ref.~\cite{Caola:2015psa}. The contribution of top quark 
loops to $gg \to ZZ + g$  amplitudes is not known analytically; 
we obtain it as an expansion in $1/m_t$ as described below. Alternatively, these amplitudes can be
obtained from one-loop providers~\cite{Cascioli:2011va,Hirschi:2011pa,Actis:2012qn,Cullen:2014yla,Hirschi:2015iia}. 
For our studies, we often used the {\tt OpenLoops} program~\cite{Cascioli:2011va}
as a cross-check of our implementation. 

Since we allow for off-shell production of the $Z$-bosons, 
we also include single-resonant amplitudes. 
Amplitudes 
$gg \to Z^* \to 4l$ receive contributions from massless and massive 
triangle diagrams  which vanish  at 
any loop order  provided that the gluons are on-shell~\cite{Hagiwara:1990dx,Campbell:2007ev}.
For this reason, we only need to consider single-resonant amplitudes with an emitted gluon 
$ gg \to g Z^* \to g + 4l $ shown 
in Fig.~\ref{fig:FeynDiagZZNLO} (f); 
the analytic expressions for amplitudes that contain both  massless and 
massive loops are given in Ref.~\cite{vanderBij:1988ac}.  We checked our implementation of single-resonant 
amplitudes  against the {\tt OpenLoops} program~\cite{Cascioli:2011va} and found good agreement. 

As previously mentioned, the top quark contribution to the  
two-loop amplitude for $gg \to ZZ$ prompt production  is intractable at present.
In order to get around this, we compute this amplitude in a heavy-top expansion, keeping terms up 
to ${\cal O}(m_t^{-8})$.
The calculation  employs the standard procedures of the large mass expansion 
(see e.g. Ref.~\cite{Smirnov:2002pj})  that allows one 
to express all contributing diagrams through a linear combination of 
vacuum bubble integrals and one-loop three-point functions with massless internal lines, which
can be easily computed.  

We also include massless and massive double-triangle diagrams  
Fig.~\ref{fig:FeynDiagZZNLO} (e)
that are anomalous and are required  
to simultaneously account for bottom and top quark contributions. The analytic 
results for these triangle  diagrams can be found in 
 Refs.~\cite{Hagiwara:1990dx,Campbell:2007ev}. 
These diagrams feature a highly off-shell $t$-channel gluon, and consequently only contribute to  final 
results at the  level of just a few per mill.

The last  amplitude that we need to consider is the top quark contribution 
to the real emission amplitude for $gg \to ZZ+g$ prompt production.   This one-loop amplitude is not 
known in a closed analytic form. We computed  it in the same way as the two-loop virtual 
amplitude, by expanding in the inverse top quark mass. Similar to the virtual correction, 
the expansion can be carried out at the level of the amplitude keeping the full dependence 
on the off-shellness of the $Z$-bosons and allowing for their subsequent decay into a lepton pair. 
Below we  discuss the conditions under which this  expansion is valid. For now, it suffices to say that 
squares of the top-quark induced amplitudes for $gg \to ZZ+g \to 4l+g$ 
have been checked against the {\tt OpenLoops} ~\cite{Cascioli:2011va} for 
a number of kinematic point with an unphysically heavy top quark 
mass, where the $1/m_t$ expansion is expected to work well. Upon doing that, good   agreement 
at the level of $10^{-5}-10^{-6}$ was found. Strictly speaking, one could have used
the results 
from {\tt OpenLoops} or other one-loop providers
to avoid the $1/m_t$ expansion 
for real emission diagrams. However this is not necessary since, as we will show in the following, the $1/m_t$ expansion
works well in the kinematics region we are interested in.

\begin{figure}[t]
\includegraphics[scale=0.6]{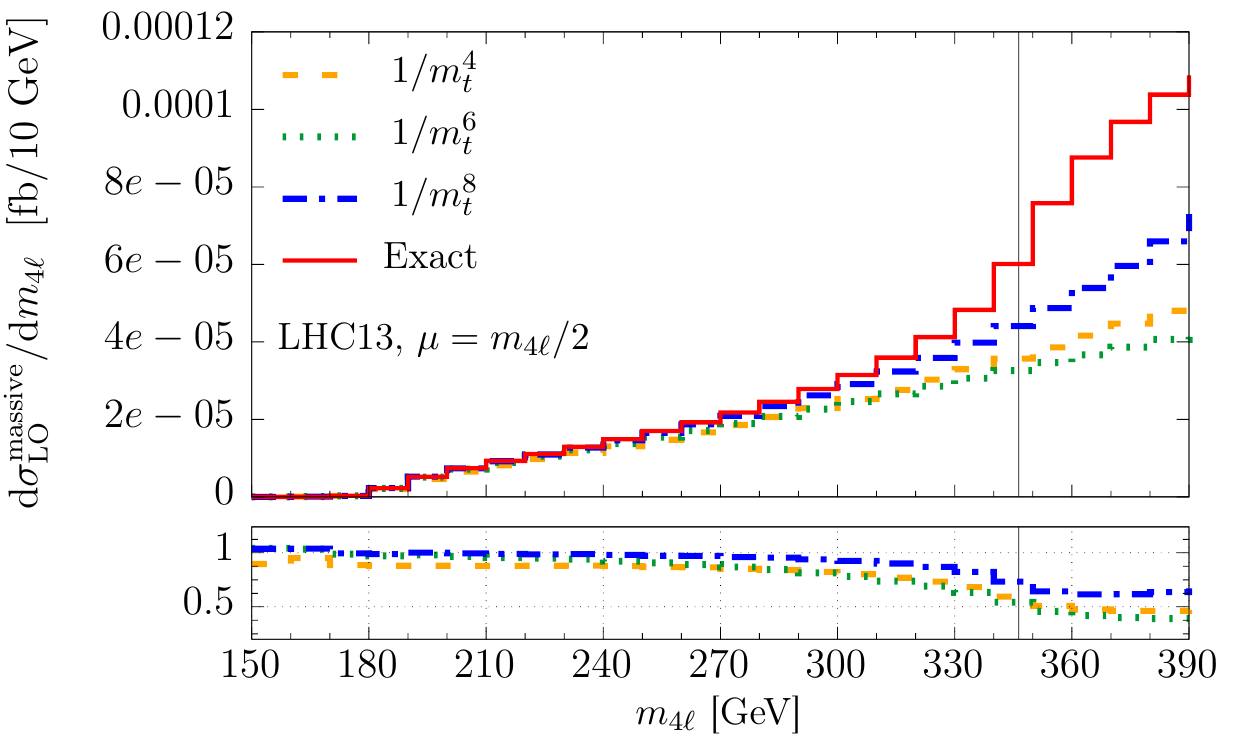}
\includegraphics[scale=0.6]{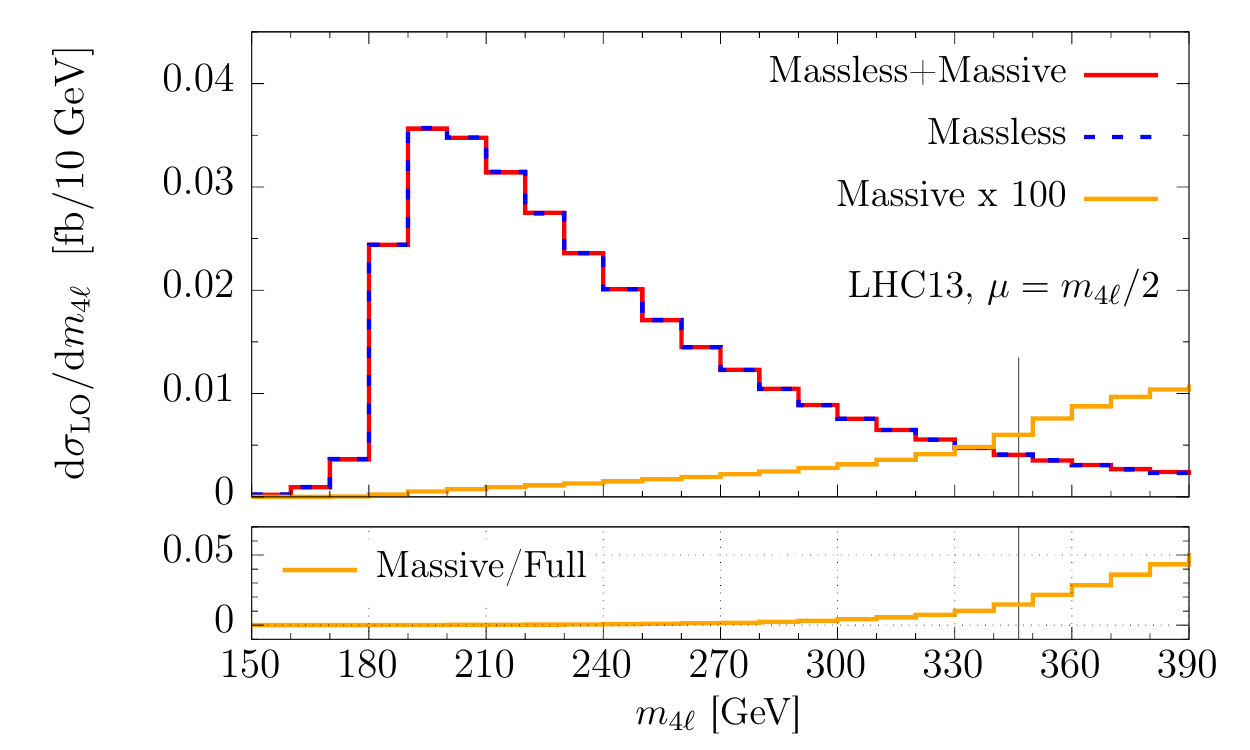}\
\caption{LO invariant mass of the four lepton system at the $13$ TeV LHC, background only.
In both plots, the vertical line marks the top threshold.
Left, upper panel: results using a massive loop only, with the amplitude evaluated in the heavy-top expansion up 
to various orders in $1/m_t$, compared to the exact mass dependence. The lower panel shows the ratio of
the various $1/m_t$ approximations to the exact result.
Right, upper panel: distribution 
using both massless and massive loops compared to massless-only and (exact) massive-only. Note that the latter is multiplied by 100.
The lower panel shows the ratio of the (exact) massive contribution to the full result.
\label{fig:bkgdLO_expcomp}}
\end{figure}

We do not consider contributions to $Z$-pair production caused by $qg$ fusion, $qg \to ZZ+q$. 
The contribution of these processes  to the interference is expected to be several times 
smaller than interference effects from the gluon fusion~\cite{Campbell:2014gua}. Moreover,
it is not possible to disentangle these contributions 
from other $qg$ or $\bar{q}g$ contributions to $ZZ$ production that appear already 
as $\mathcal{O}(\alpha_s^3)$ corrections to the main production mechanism $q \bar q \to ZZ$. 
With this choice, our result is contaminated by non-canceling factorization scale terms which
are however suppressed by the ratio of quark to gluon luminosities, i.e. comparable to other
terms we neglect in the full $\mathcal{O}(\alpha_s^3)$ $q\bar q\to ZZ$ computation. We prefer
not to include these terms to avoid artificially small factorization scale uncertainties, although
a proper study of their effect is beyond the scope of this work.

We will now discuss a number of checks that validate the implementation of all the amplitudes 
in our numerical code and the validation of the approximate treatment of top quark mass effects. 
The implementation of all the various amplitudes in our code was checked extensively by comparing a
large number of leading order kinematic distributions with {\tt MCFM}~\cite{Campbell:2013una} and 
by comparing the various one-loop amplitudes against {\tt OpenLoops} ~\cite{Cascioli:2011va}. 
As we already mentioned, for these checks we often take the top quark mass to have an unphysically 
large value, to ensure that the mass expansion of the amplitudes converges.  Nevertheless, 
these checks of the implementation 
still leave as an open question  whether the $1/m_t$ expansion of physical 
cross sections for $ZZ$ production in gluon fusion actually converges.

To investigate  this issue, we begin at LO, where we can perform a comparison  of exact and expanded  
in $1/m_t$ contributions to  prompt production of $Z$-pairs.  Such a comparison 
is shown in the left pane in Fig.~\ref{fig:bkgdLO_expcomp}. We see that the $1/m_t$ expansion works decently all the  
way up to $\m4l \lesssim 320~{\rm GeV}$; after that the exact and expanded result show significant differences. 
We now combine contributions of leading order massless and massive loops to $gg \to ZZ$ prompt production 
and show the ensuing  ${\rm d}\sigma_p/{\rm d}\m4l$ in the right pane of Fig.~\ref{fig:bkgdLO_expcomp}.
The result  clearly demonstrates that contributions of massless loops dominate so strongly that any issues 
with expansions in $1/m_t$ around the top quark threshold, visible in the left pane, 
become unobservable. Indeed, below the top threshold the deviations between exact and approximate results
seen in Fig.~\ref{fig:bkgdLO_expcomp} only affect the total result at the sub-percent level.

\begin{figure}[t]
\includegraphics[scale=0.7]{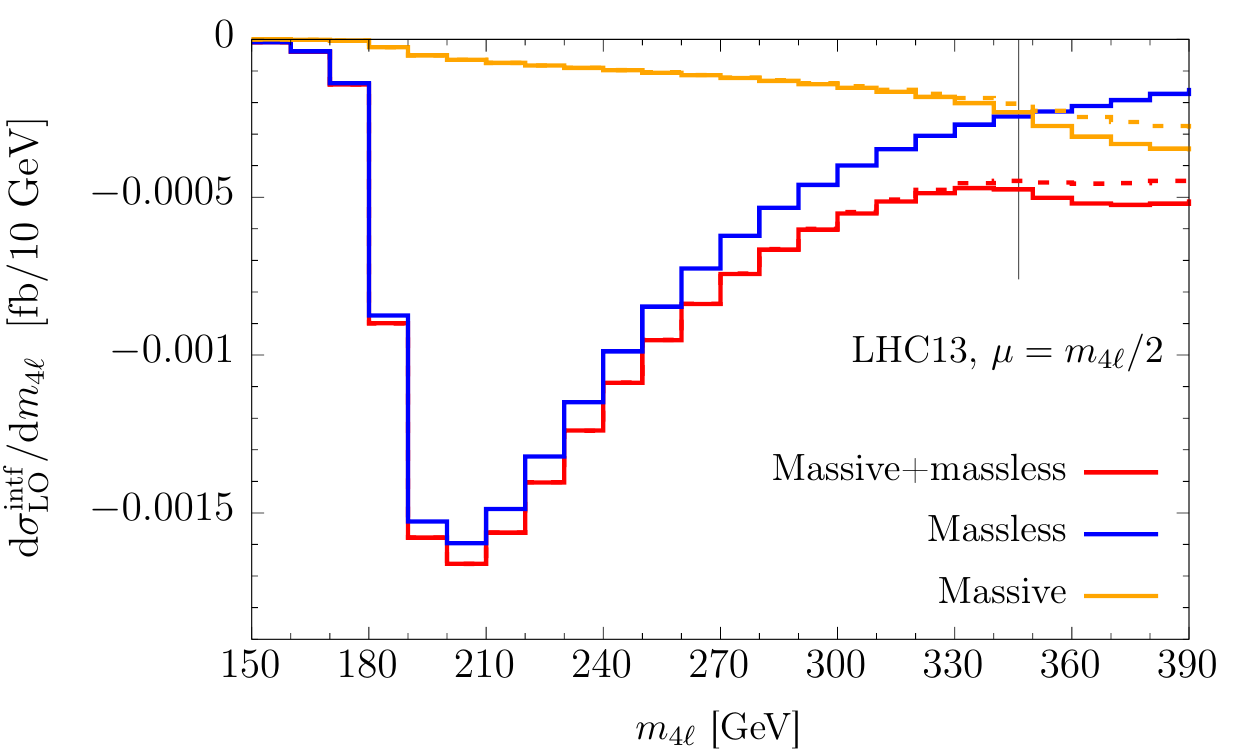}
\caption{LO results for signal/background interference at the 13 TeV LHC. Both the full result 
as well as massless/massive-only 
contributions are shown. Solid line: exact result. Dashed line: 
$1/m_t$ expansion, including up to $1/m_t^8$ terms. 
The vertical line marks the top threshold.\label{fig:intfLO_expcomp}}
\end{figure}

The situation is however different if one considers the interference between signal and background.
Indeed, it is expected on general grounds that top quark contributions to the interference 
play a much more important role,  because for $ \m4l \ge 2m_Z$, the off-shell 
Higgs boson decays preferentially to longitudinal $Z$-bosons. In turn, the longitudinal $Z$-bosons 
have stronger couplings to top quark loops than to massless loops; as a result the contribution of 
top quark loops  is more prominent in the interference than in the background cross section.  
These expectations are confirmed in Fig.~\ref{fig:intfLO_expcomp} where  we show 
the interference contribution to the $\m4l$ invariant mass distribution. 
Although the qualitative behavior of massless and massive contributions to the full result 
is similar to the pure background case -- massless/massive contribution decreasing/increasing with $\m4l$ -- 
the impact of massive amplitudes is quite sizable.
At the top quark threshold $\m4l \sim 2 m_t$, the two contributions become comparable.  
At this value of $\m4l$, the differences between exact and $1/m_t$-expanded results start to appear. 
Still, it follows from Fig.~\ref{fig:intfLO_expcomp}, that the error associated with using 
the $1/m_t$ expansion for the interference  
is a few percent even at the high end of the expansion   region which, 
as we will see,  is smaller than other sources of uncertainty such as uncalculated higher order corrections.
We therefore conclude that we can use the heavy top quark mass expansions to study the interference 
in $gg \to ZZ$ provided that  we restrict ourselves to $\m4l \le  2 m_t$.

\begin{figure}[t]
\includegraphics[scale=0.7]{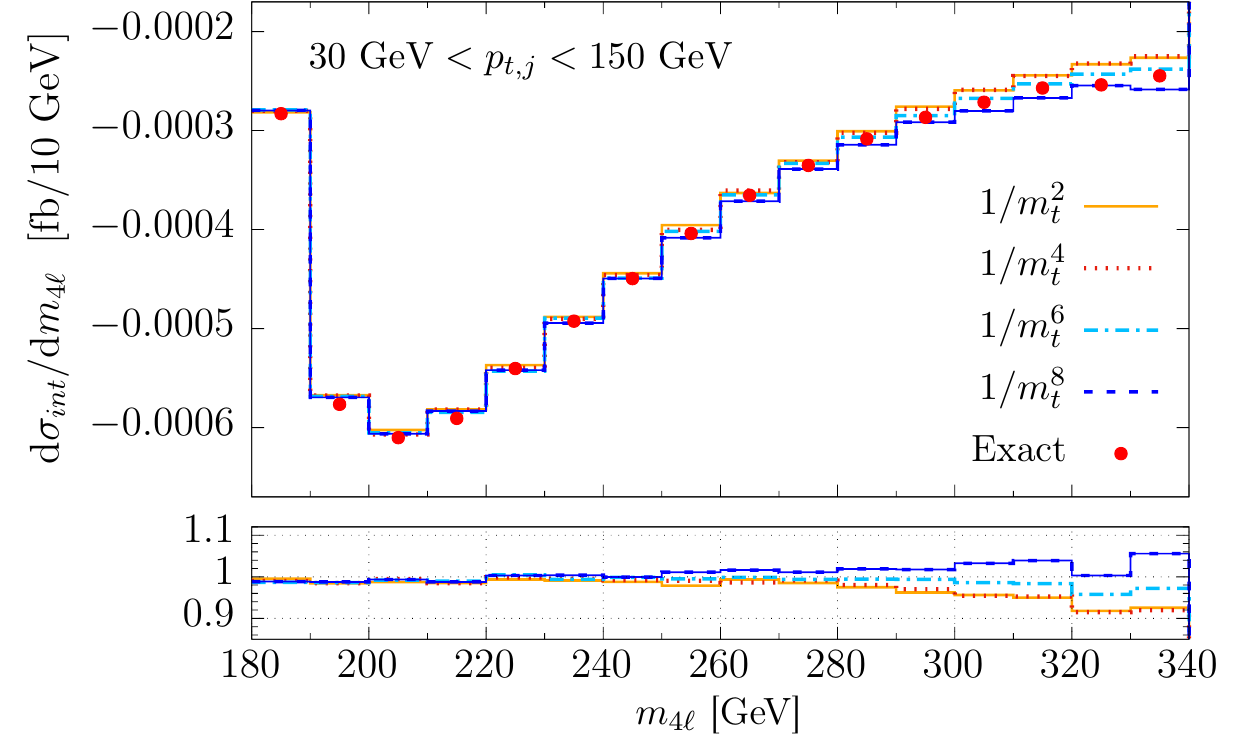}
\caption{Interference pattern in $gg \to ZZ + {\rm jet}$ between 
the Higgs signal and the prompt production, at the $\sqrt{s}=13$ TeV LHC for a scale $\mu=m_{ZZ}/2$, including both massless and massive contributions.
A comparison of expanded in $1/m_t$ and exact 
results are shown, with the latter taken from Ref.~\cite{Campbell:2014gua}.
\label{fig:intfreal_expcomp}}
\end{figure}

Since the kinematic features  of the virtual corrections are identical 
to those of leading order amplitudes, the $1/m_t$ expansion of the two-loop amplitude is expected to be  
valid for $\m4l < 2 m_t$ as well.
This is not necessarily the case for the real emission contributions, 
since a hard gluon emission can resolve the top loop even if $\m4l < 2 m_t$. 
In order to understand the effect of a hard jet on the 
$1/m_t$ 
expansion, we 
compare our results with those 
of  Ref.~\cite{Campbell:2014gua},
where the interference effects are calculated in the presence of hard jet.
We note that while this calculation includes the full mass dependence in the amplitudes, 
it only considers on-shell $Z$-bosons whose decays are modeled by 
multiplying the result of the calculation by the $Z\to \myell \myell$ branching ratios. For the sake 
of comparison, we can circumvent these differences by keeping the $Z$-bosons on-shell and integrating over the phase space of the 
produced leptons. We then compare the $\m4l$ distribution of the interference contribution 
in $gg \to ZZ + g$ above the $2 m_Z$ threshold  
with the $m_{ZZ}$ distribution of Ref.~\cite{Campbell:2014gua}. We find that, in order to have 
an agreement between expanded and exact results, we need to introduce an upper cut on the 
transverse momentum of a jet $p_{\perp  j}^{\rm max}$. We need to choose $p_{\perp  j}^{\rm max}$
as large as possible and, at the same time, attempt to maintain  the convergence of the $1/m_t$ expansion. 
We have found, empirically, that a cut as large as  $p_{j,\perp}^{\rm max} < 150~{\rm GeV}$ allows us 
to obtain good agreement   with the calculation of Ref.~\cite{Campbell:2014gua} all the way up to the 
top production threshold (see Fig.~\ref{fig:intfreal_expcomp}), while only excluding about 8\% of hard jet events.
We will use this cut  when we study 
the LHC phenomenology in the next section. We conclude this section by stressing that in this
paper we are mostly interested in genuine NLO corrections to the $gg\to 4l$ process. As a consequence,
this relatively hard upper cut on jet emission is not particularly relevant for us, since in the region $p_{\perp j} > p_{\perp,j}^{\rm max}$
our computation is only LO and if desired the result in this region can be obtained using automatic one-loop
frameworks, see e.g.~\cite{Cascioli:2013gfa,Hirschi:2015iia}.

\subsection{LHC phenomenology} \label{sec:ZZpheno}

In this section, we present the calculation   of the NLO QCD corrections 
to $gg \to ZZ$ production at the $\sqrt{s}=13$ TeV LHC,  including off-shell Higgs and the interference 
effects. For background processes, $\gamma^*$-mediated processes are included as well.
 We employ  the following parameters in our  computation 
\begin{equation}
\begin{split}
m_Z =& 91.1876~\mrm{GeV},  \hspace{1.5cm} m_W = 80.398~\mrm{GeV}, \\
\Gamma_Z =& 2.4952~\mrm{GeV},  \hspace{1.5cm} \Gamma_W = 2.1054~\mrm{GeV}, \\
m_t=&173.2~\mrm{GeV}, \hspace{1.5cm} G_F=1.16639 \times 10^{-5}~\mrm{GeV}^{-2}, \\
g_w^2=& 4\sqrt{2} m_W^2 G_F, \hspace{1.5cm} \sin^2\theta_W = 0.2226459. 
\end{split}
\end{equation}
We use the  bottom quark mass $m_b=4.5~\mrm{GeV}$ when evaluating 
the amplitude for Higgs-mediated processes. However,  we take $m_b$ to be massless 
when computing  the amplitudes for the prompt production process $gg \to ZZ$. 
We use LO and NLO NNPDF3.0 parton distribution functions~\cite{Ball:2014uwa}
to obtain  leading and next-to-leading order  results,  respectively.
We use dynamical renormalization and factorization scales with 
the central value  $\mu_0 = \m4l/2$,  
and vary it by a factor of two in either direction to estimate the scale dependence 
of the final result.

Apart from the restrictions on $\m4l$ and the jet transverse momentum discussed in the previous section, 
we only impose cuts on the invariant mass of the lepton pairs, $60~\mrm{GeV} < m_{\myell \myell} < 120~\mrm{GeV}$ 
to isolate the $Z$ resonance and suppress $\gamma^*$ contributions.

We start by  considering  the  $gg \to ZZ \to e^+ e^- \mu^+ \mu^-$  production in the four-lepton 
invariant mass interval $150~\mrm{GeV} < \m4l < 340~\mrm{GeV}$. 
The lower boundary  separates Higgs off-shell events  from Higgs  on-shell events; the upper boundary 
is imposed to ensure the validity of the $1/m_t$ expansion. 

We begin by presenting the results for the cross sections for $gg \to ZZ \to e^+ e^- \mu^+ \mu^-$ 
in the interval of four-lepton invariant masses described above. We show results for 
the signal, the background, the interference and the full cross section. We find the following 
results at leading and next-to-leading orders in perturbative QCD 
\begin{equation}
\begin{split}
& \sigmasiglLO=0.043^{+0.012}_{-0.009}~\mrm{fb},\;\;\; \;\;\;\sigmasiglNLO=0.074^{+0.008}_{-0.008}~\mrm{fb} \\
& \sigmabkgdLO=2.90^{+0.77}_{-0.58}~\mrm{fb}, \;\;\;\;\;\;\;\; \sigmabkgdNLO=4.49^{+0.34}_{-0.38}~\mrm{fb} \\
& \sigmaintfLO=-0.154^{+0.031}_{-0.04}~\mrm{fb},\;\;\;\;\; \sigmaintfNLO=-0.287^{+0.031}_{-0.037}~\mrm{fb}  \\
& \sigmafullLO=2.79^{+0.74 }_{-0.56}~\mrm{fb},\;\;\;\;\;\;\;\;\sigmafullNLO=4.27^{+0.32}_{-0.35}~\mrm{fb},
\end{split}
\end{equation}
where the sub- and superscripts indicate the scale variation.
The interference is destructive, as implied  by the unitarity arguments, 
despite the fact that these cross sections refer to the production of four leptons with 
invariant masses that are far below the values for which the unitarity arguments are valid. 
 Negative interference implies that the
physical cross section is smaller than the sum of the signal and background cross sections 
by about 5\%.
We also note that the absolute value of the interference is 3-4 times larger than the signal, 
but is still more than an order of magnitude smaller than the irreducible $gg$  background.
Consequently, extracting the signal and observing the effect of the interference in this range of four-lepton 
invariant masses will be challenging, assuming the Higgs couplings to vector bosons and gluons 
are close to what is expected in the Standard Model. 

We observe that the 
 NLO QCD corrections are largest for the signal cross section and smallest 
for the background. The corresponding $K$-factors\footnote{We define 
the $K$-factor as the ratio of NLO corrected cross section at a particular 
scale to the leading order cross section at the central scale.} are  
$K_{\rm signal} = 1.72$ and $K_{\rm bkgd} = 1.55$ for the central scale choice.
It is interesting to note that the $K$-factor for the interference, $K_{\rm intf} = 1.65$, 
is very close to the geometric mean of these results $K_{\rm intf} \approx \sqrt{K_{\rm bkgd} K_{\rm signal} }$, 
as was assumed in experimental analyses aimed at constraining the  Higgs boson width 
\cite{Khachatryan:2014iha,Khachatryan:2015mma}.
The scale uncertainty of the leading order cross section 
 is in the range of twenty to thirty percent; 
the NLO cross sections are outside the scale uncertainty of the leading order result. 
At NLO, the relative scale uncertainty decreases by about a factor of two and becomes 
close to ten percent. 

\begin{figure}[t]
\includegraphics[scale=0.6]{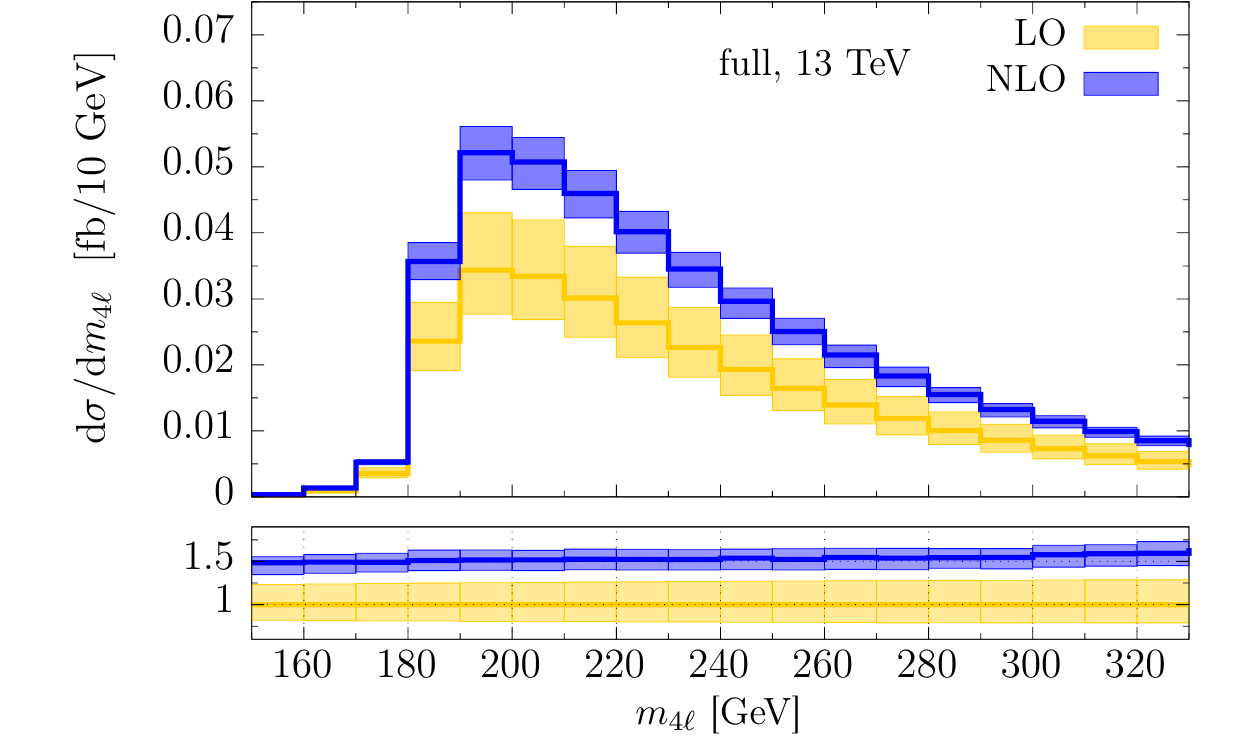}
\includegraphics[scale=0.6]{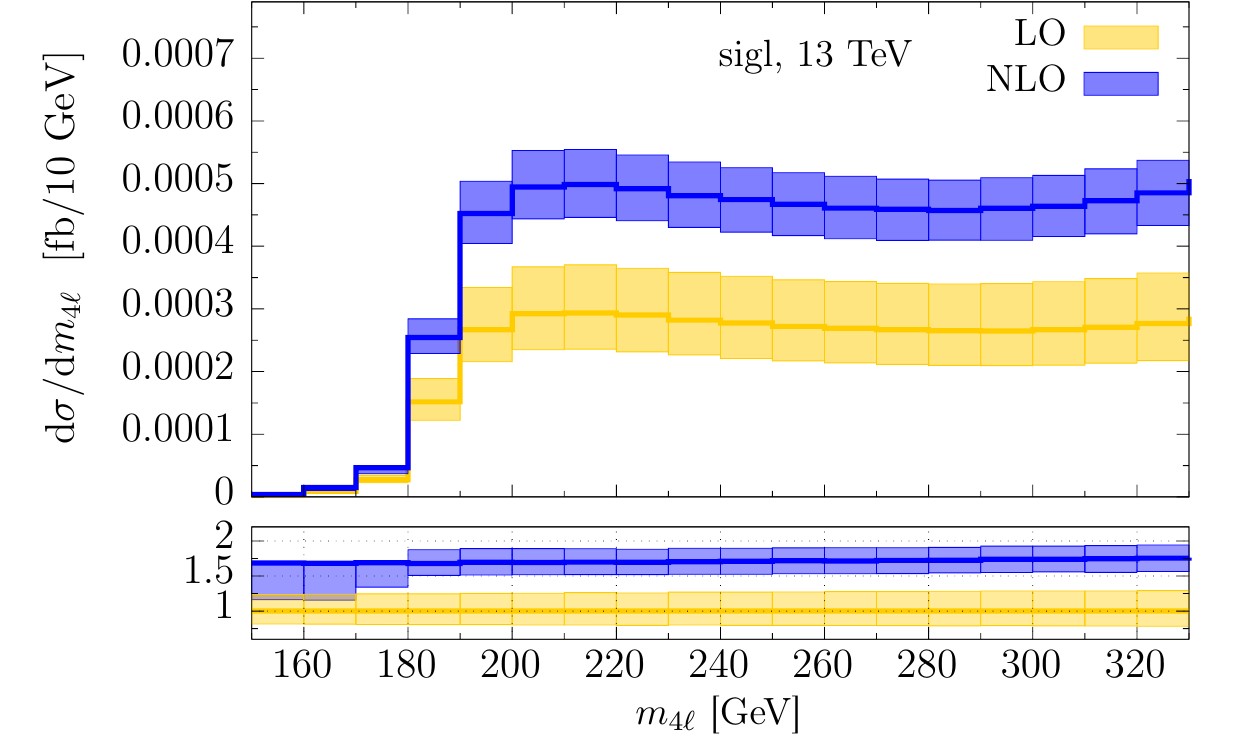}
\vskip0.8cm
\includegraphics[scale=0.6]{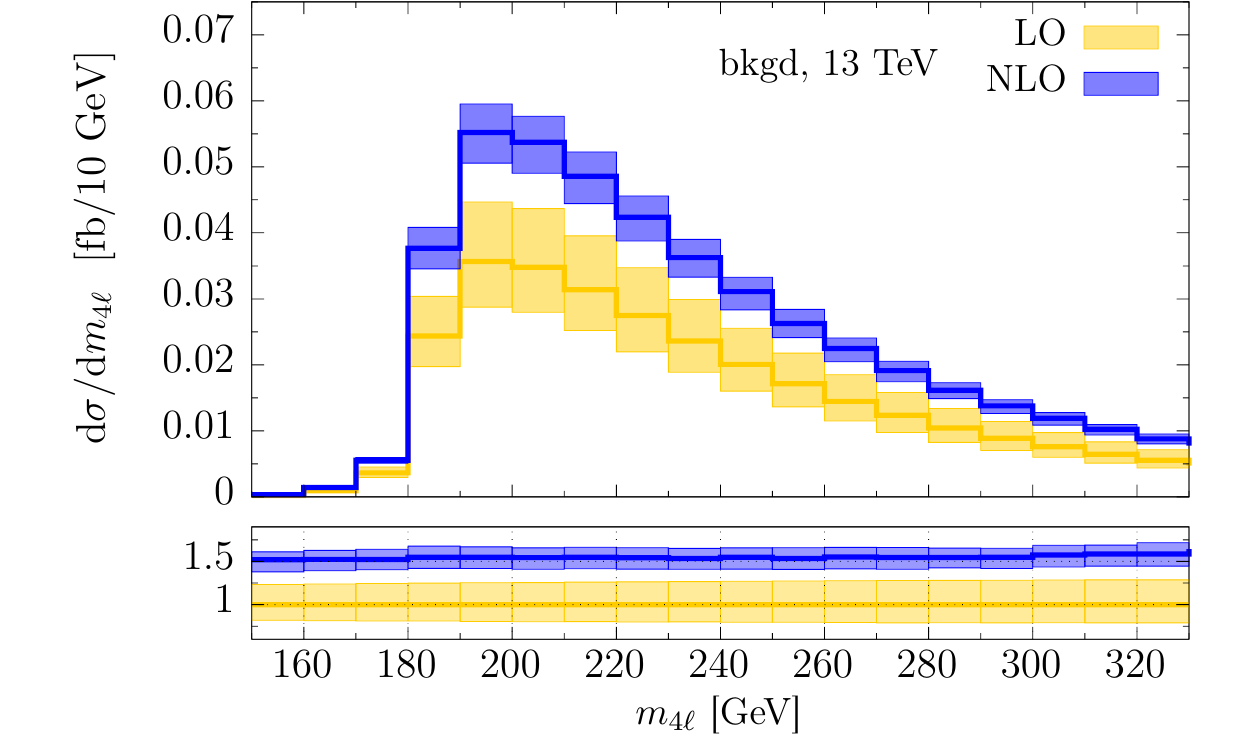}
\includegraphics[scale=0.6]{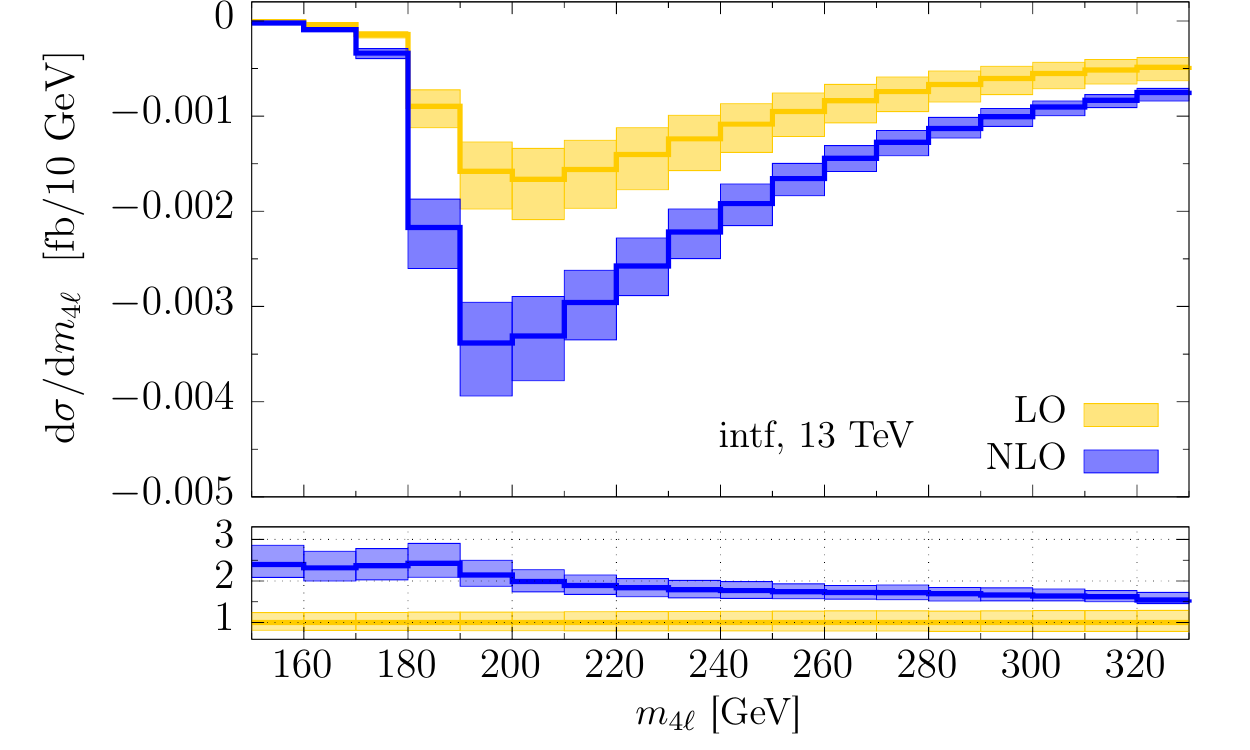}
\caption{Four-lepton invariant mass distributions in $gg \to ZZ$ processes at the $13$ TeV LHC.
The full result is shown as well as contributions of signal, background and interference separately.
LO results are shown in yellow,  NLO results are shown in blue, and scale variation is shown for $\m4l/4 < \mu < \m4l$ with a central scale $\mu=\m4l/2$.
The lower pane shows the $K$-factors.
\label{fig:ZZ_m4l_scale}}
\end{figure}

We continue with  the discussion of differential distributions in the invariant mass 
of four leptons, $\m4l$.  In Fig.~\ref{fig:ZZ_m4l_scale} we show separately the 
distributions for the signal, the background, the interference and the total yield of four leptons 
in gluon fusion. The lower panes show the corresponding $K$-factors, in dependence of $\m4l$.  
We note that $K$-factors for the signal and the background distributions are relatively flat, 
with a slight  increase with $\m4l$.  The situation with the interference is different. 
In this case, the $K$-factor around the $2m_Z$ threshold is large, 
$K_{\rm intf} \approx 2.5$ for $\m4l \lesssim 2 m_Z$.  As the invariant mass increases, the interference 
$K$-factor decreases rapidly and flattens out, 
reaching the  value $K_{\rm intf} \approx  1.5$ at $\m4l = 2m_t$.   Hence, at around $\m4l \sim 2 m_t$, 
values of the interference, signal and background $K$-factors become very similar and, practically, 
independent of the value of the invariant mass $m_{4\myell}$. 
Thus,  we find that the impact of NLO QCD corrections on the interference $K$-factor 
can be approximated by the geometric mean of the signal and the background $K$-factors when 
the interference is integrated over the full kinematic range of four-lepton masses, as well as 
at higher values of the invariant masses where $K_{\rm signal} \approx K_{\rm bkgd} \approx K_{\rm intf}$.
However, this is not the case close to $2 m_Z$ threshold, where the behavior of the 
interference $K$-factor is different from either the signal or background $K$-factors.

\begin{figure}[t]
\includegraphics[scale=0.6]{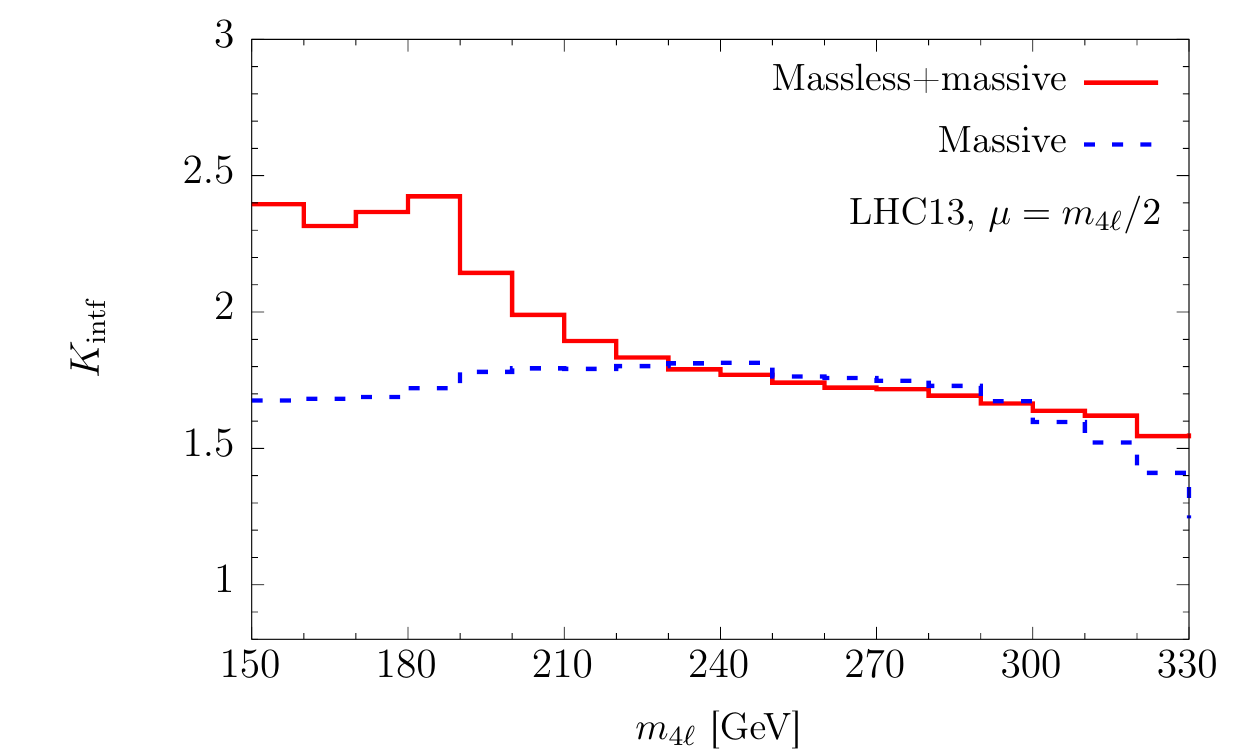}
\caption{Comparison of full (massive+massless) and massive only interference $K$-factors as a function of $m_{4\myell}$
at the 13 TeV LHC.
\label{fig:ZZ_m4l_KFmm}}
\end{figure}

Finally, we compare the behavior of the NLO corrections to the interference arising from massive prompt production amplitudes to the full interference result which arises from both massless and massive loops.
Such a comparison is shown in Fig.~\ref{fig:ZZ_m4l_KFmm}. 
We have already seen that the massless contribution strongly dominates the interference at around 
$m_{4\myell} \sim 2 m_Z$ and, as seen from the behavior of the full result, drives a rapid increase in the $K$-factor close to $m_{4\myell} \sim 2 m_Z$. 
In contrast, $K_{\rm intf}$ for the massive loops remains flat for $\m4l \lesssim 280~\mrm{GeV}$, at which point it begins to decrease.
This means that $K_{\rm intf}$ for the massive loops is well approximated by the geometric mean $\sqrt{K_{\rm bkgd} K_{\rm signal} }$ across the full range of $\m4l$ that we consider.

\section{$WW$ production} \label{sec:WW}

In this section, we discuss the production of $W$-boson pairs  in gluon fusion, including 
interference effects. 
Such interference effects have previously been studied at LO in Refs.~\cite{Campbell:2011cu,Campbell:2013wga}.
Unlike in $ZZ$ production, the invariant mass of the off-shell Higgs is not observable, because of neutrinos in the final state. 
Nevertheless, it is possible to use 
the transverse mass of the $WW$ system  to probe the off-shell Higgs physics.

The set up of our calculation is similar to the case of $Z$-boson pair production described  in the previous 
section. The principal 
difference between the two cases is that for $W$-boson pair production, we do not include the contribution 
of the third generation of quarks when computing QCD radiative corrections. 
It is known that this contribution amounts to approximately 10\% of the gluonic $WW$ cross section at LO~\cite{Campbell:2011cu,Kauer:2013cga}.
As mentioned in the Introduction, this omission is due to the complexity of performing a mass expansion with both top and bottom quarks in the loop.
Therefore, our results for $gg \to WW$ are necessarily incomplete 
but  
they, at least, give partial information about 
radiative effects in the case of the $WW$ production in gluon fusion.

The amplitudes for  $gg \to WW$  production 
are assembled along the lines described in Section~\ref{sec:ZZ}. 
We consider leptonic decays of the $W$-bosons, $gg \to WW \to \nu_e e^+ \mu^- \bar{\nu}_{\mu}$, and consistently 
include  the required single-resonance contributions;  this  allows us to describe the $W$-pair  production 
for a broad range of invariant masses both below and above the $2m_W$-threshold. 

\begin{figure}[t]
\includegraphics[scale=0.6]{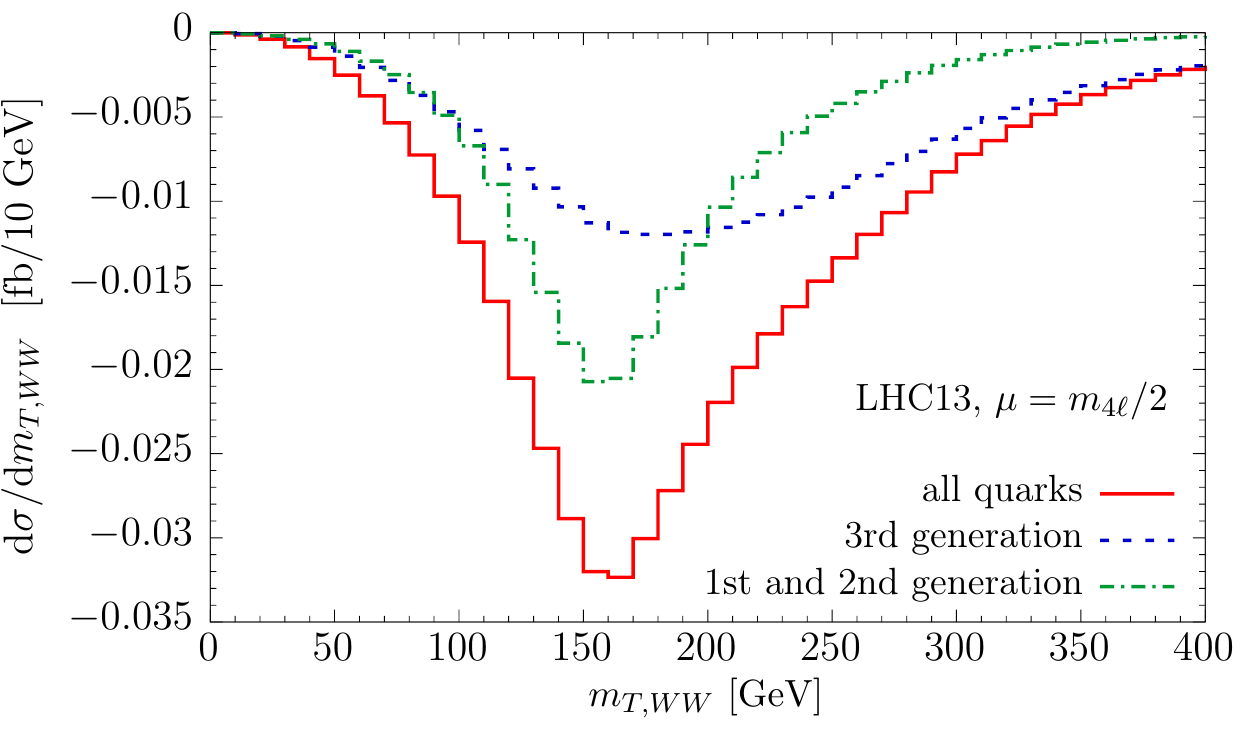}
\caption{Contributions of the first two generations and the third generation to the interference 
in $gg \to W^+W^-$  at leading order in perturbative QCD.  
\label{fig:ZZ_mTWW_intflo}}
\end{figure}

We present results for the $\sqrt{s}=13$ TeV LHC, using the same parameters, scales 
and parton distribution functions as in the previous section.
Since we do not use an expansion in $1/m_t$, we no longer require the cut $p_{\perp j} < p_{\perp  j}^{\rm max}$, and we remove this cut from our analysis.
We also do not impose any cuts on the final state leptons, so that the results shown in this Section are fully inclusive.
We stress, however, that our computation can accommodate any cut on final state leptons, missing energy and jet.

We begin with the discussion of the interplay between contributions of the 
 third and the first two generations to the interference at leading order. The results 
are shown in Fig.~\ref{fig:ZZ_mTWW_intflo} where a comparison is made in dependence of the 
transverse mass $m_{T, W^+W^-}$. The transverse mass is defined
\be
m_{\rm T,WW} = \sqrt{2  E_{\perp,{\rm miss}} p_{\rm T,\myell \myell} (1-\cos \tilde \phi)}
\ee
where $p_{\perp, \myell\myell}$ is the transverse momentum of the lepton pair, $E_{\perp,{\rm miss}}$ is the missing energy, 
and $\tilde \phi$ is the azimuthal angle between the direction of the missing energy and the $\myell^+ \myell^-$ system.
It is apparent 
from Fig.~\ref{fig:ZZ_mTWW_intflo} that, unlike the situation for the $gg \to WW$ cross section, the massless contributions to the interference do not 
dominate for any value of $m_{T,W^+W^-}$ \cite{Campbell:2011cu}. 
In fact, the two first generations and the third generation 
give, roughly, comparable contributions to the interference, for $m_{T,W^+W^-} \le 200~{\rm GeV}$; 
for higher values of the transverse mass, the contribution of the third generation dominates.  

\begin{figure}[t]
\includegraphics[scale=0.6]{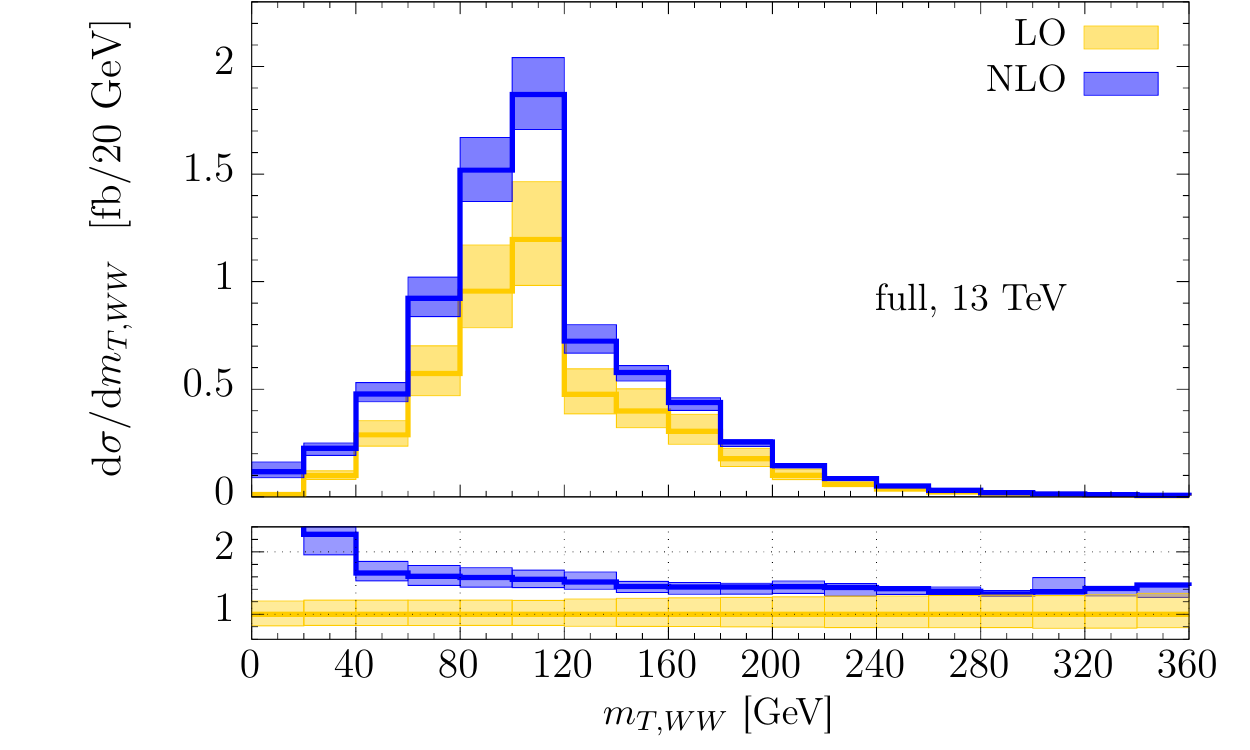}
\includegraphics[scale=0.6]{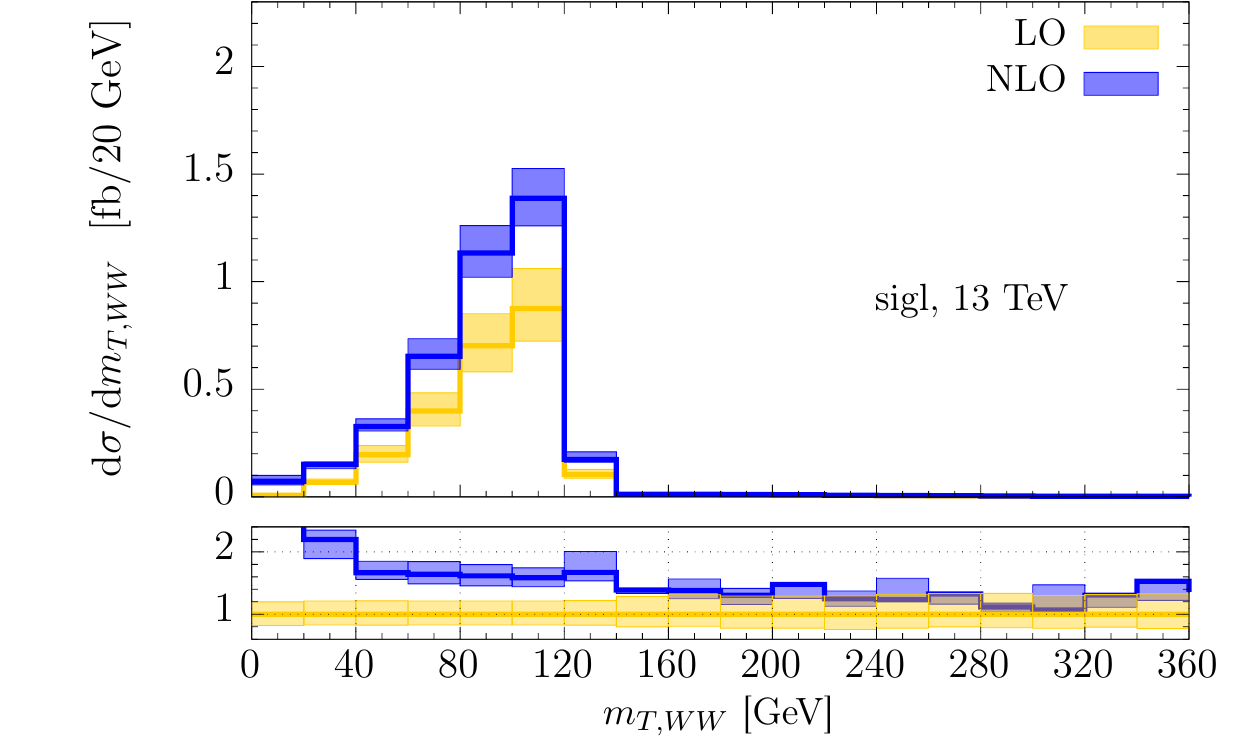}
\vskip0.5cm
\includegraphics[scale=0.6]{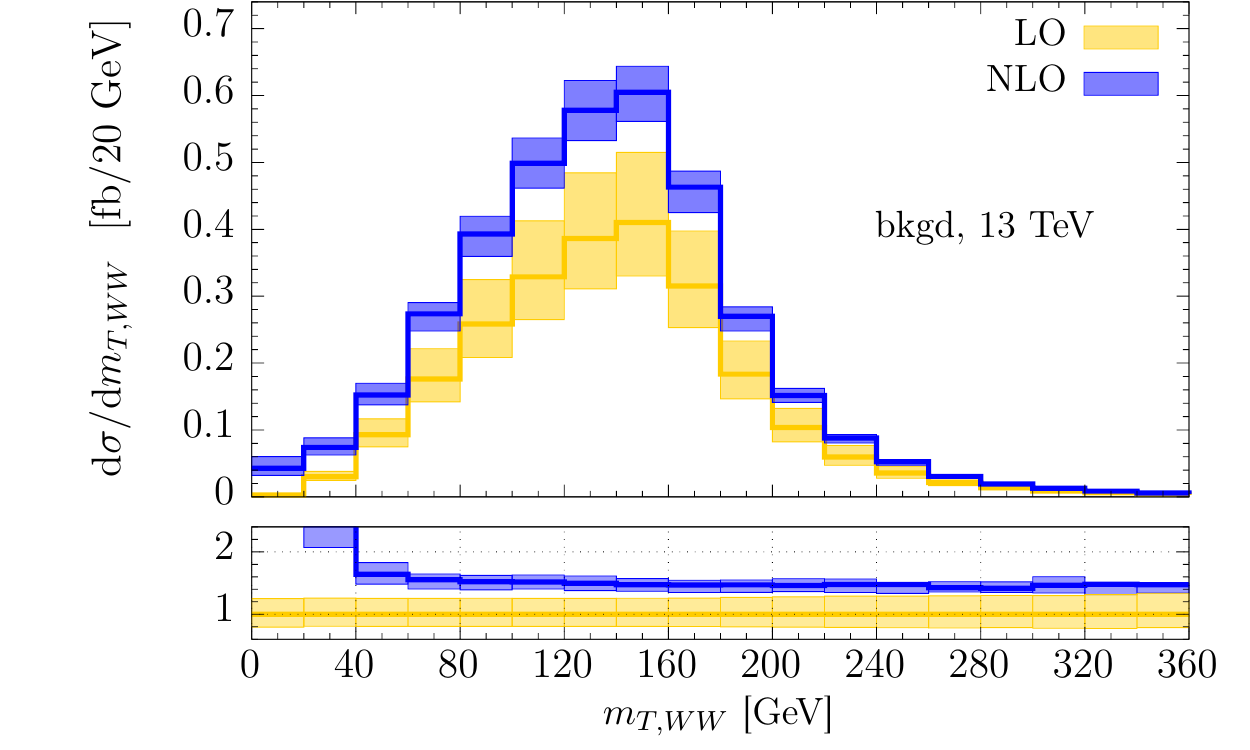}
\includegraphics[scale=0.6]{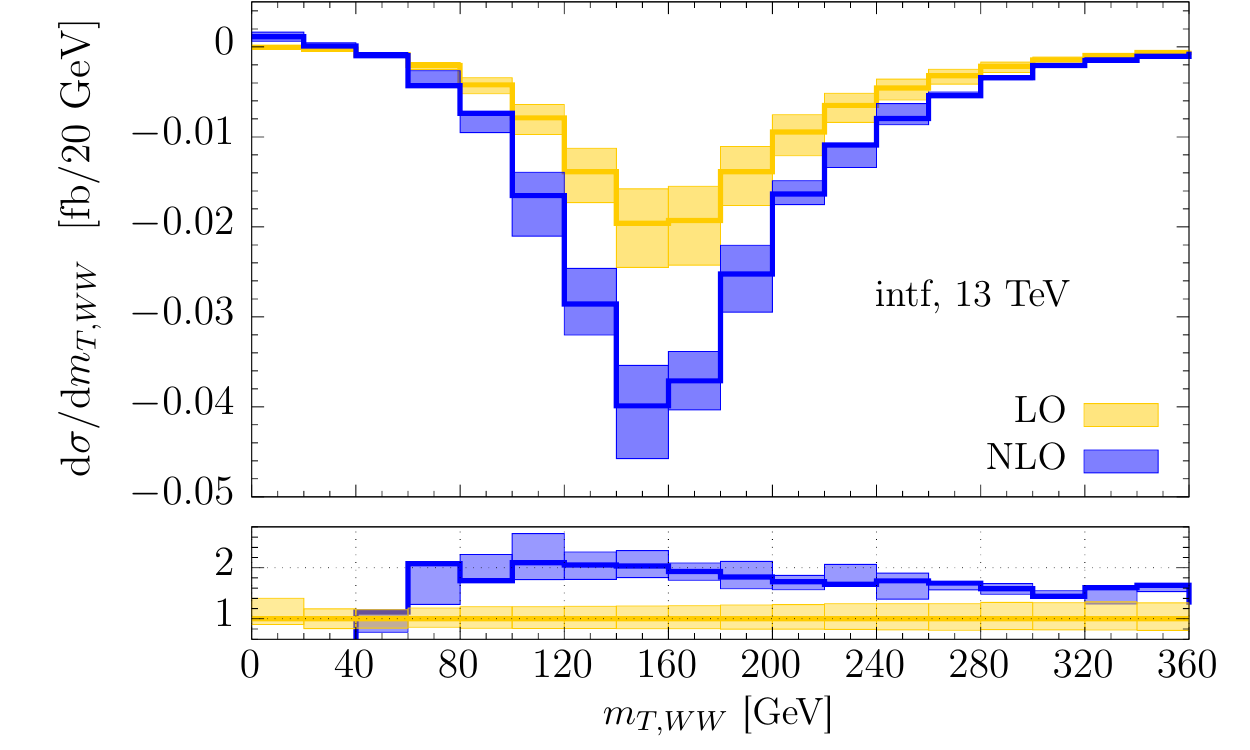}
\caption{Transverse mass $m_{T,WW}$ distributions in $gg \to WW$ process at the $13$ TeV LHC.
The full result is shown as well as contributions of signal, background and interference separately. Only contributions from the first two quark generations to the prompt production amplitudes are included.
LO results are shown in yellow,  NLO results are shown in blue, and scale variation is shown for $\m4l/4 < \mu < \m4l$ with a central scale $\mu=\m4l/2$. 
The lower pane shows the $K$-factors.
\label{fig:WW_mtww_scale}}
\end{figure}

We now turn to the impact of NLO corrections. We consider only contributions from the \textit{first two quark generations} to the prompt production amplitudes at both LO and NLO, so as to treat the results on an equal footing. At $\sqrt{s}=13$ TeV, the cross sections are
\begin{equation}
\begin{split}
& \sigmasiglLO=48.3^{+10.4}_{-8.4}~\mrm{fb},\;\;\; \;\;\;\sigmasiglNLO=81.0^{+10.5}_{-8.2}~\mrm{fb} \\
& \sigmabkgdLO=49.0^{+12.8}_{-9.7}~\mrm{fb}, \;\;\;\;\;\;\;\; \sigmabkgdNLO=74.7^{+5.5}_{-6.2}~\mrm{fb} \\
& \sigmaintfLO=-2.24^{+0.44}_{-0.59}~\mrm{fb},\;\;\;\;\; \sigmaintfNLO=-4.15^{+0.47}_{-0.54}~\mrm{fb}  \\
& \sigmafullLO=95.0^{+22.6}_{-17.6}~\mrm{fb},\;\;\;\;\;\;\;\;\sigmafullNLO=151.6^{+15.4}_{-13.9}~\mrm{fb}.
\end{split}
\end{equation}
Similar to  $ZZ$ production studied in the previous section, interference is destructive, although less important, reducing the full cross section by about 2\%-3\%.
In contrast to $ZZ$ production,  we do not remove the 
kinematic region corresponding to the Higgs peak, resulting 
in a signal cross section that is comparable to the background, 
and more than an order of 
magnitude greater than the interference.
 
It is well-understood how to 
 construct cuts to either suppress or 
enhance the relative contribution of the interference\footnote{See e.g. 
Ref.~\cite{Aad:2015xua} for a description of experimental 
selection criteria in off-shell studies.}  and we emphasize 
that, since our computation  includes off-shell 
effects and decays of the $W$-bosons, we can 
implement any such  cuts within our numerical code.

The NLO corrections enhance the signal and background cross sections by $K_{\rm signal}=1.68$ and $K_{\rm bkgd}=1.53$ respectively, similar to the $K$-values found for $ZZ$ production in the previous section. 
However, for the interference $K_{\rm intf}=1.85$, which is larger than the corresponding $K$-factor in $Z$-pair production, $K_{\rm intf}=1.65$. 
While the relationship between the interference $K$-factor and the geometric mean $\sqrt{K_{\rm bkgd} K_{\rm signal} }$ is no longer exact, 
the geometric mean still provides a decent approximation to $K_{\rm intf}$.

\begin{figure}[t]
\includegraphics[scale=0.7]{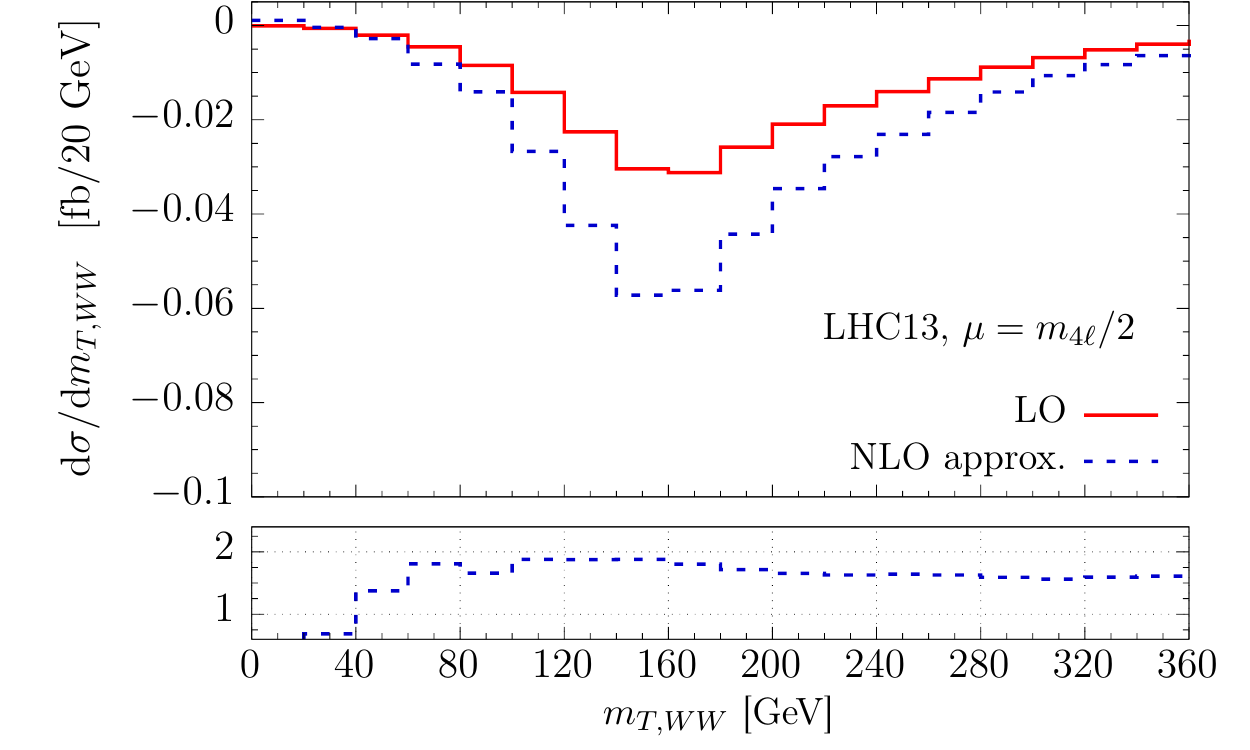}
\caption{Transverse mass $m_{T,WW}$ distribution for the interference in $gg \to WW$ at the $13$ TeV LHC. The LO result includes 
contributions from three quark generations. 
The NLO result is obtained by summing the exact result 
for the first two generations with  the LO third generation 
contribution multiplied by a constant $K$-factor 
$\sqrt{K_{\rm bkgd} K_{\rm signal} }=1.6$. \label{fig:WW_intf_mm}}
\end{figure}

We show the $m_{T,WW}$ distributions for the signal, the background, the interference, and the total yield in Fig.~\ref{fig:WW_mtww_scale}, with the $K$-factors in the lower panes.
We note that all LO distributions approach zero for low $m_{T,WW}$, leading to extremely large $K$-factors in this region. 
Apart from this,  $K_{\rm bkgd}$ is relatively flat, as is $K_{\rm signal}$ for $m_{T,WW} < 2m_W$, after which the signal is suppressed and the statistics are limited.
The $K$-factor for the interference again behaves differently, dropping from $K_{\rm intf} \approx 2$ at $m_{T,WW} \approx 60$ GeV to $K_{\rm intf} \approx 1.5$ at the high end of the distribution. 
A qualitatively similar effect was seen in the $\m4l$ distributions from $ZZ$ production, which again was ascribed to the massless contributions, while the massive contribution remained relatively flat (cf. Fig.~\ref{fig:ZZ_m4l_KFmm}). 
This observation suggests a way of estimating the impact of NLO QCD corrections to the interference including all quark flavors, by adding the NLO results displayed in Fig.~\ref{fig:WW_mtww_scale} to the LO third generation 
contribution  multiplied by the approximate 
$K$-factor $\sqrt{K_{\rm bkgd} K_{\rm signal} }=1.6$. This procedure 
results in an approximate NLO interference cross section 
$\sigma^{\rm intf}_{\rm NLO,approx.}=-8.35$ fb, to be compared with a 
LO result of $\sigmaintfLO=-4.86$ fb including all quark contributions. 
The corresponding $m_{T,WW}$ distribution is shown in Fig.~\ref{fig:WW_intf_mm}. 
Finally, we reiterate that this approximation to the full NLO interference 
can be improved by calculating the massive loops either in a $1/m_t$ 
expansion or with the full mass dependence. While the latter is at 
the limit of our current capabilities, it is the only way in which 
mass effects can be unambiguously included in all kinematic regimes.

\section{Conclusion} \label{sec:concl}

In this paper, we described the computation of 
 the NLO QCD radiative corrections to the production of four leptons
in gluon fusion, $gg \to VV\to 4l$, $V=Z/\gamma^*,W$ and discussed phenomenological 
implications for the LHC.   Our computation includes both prompt and Higgs-mediated production mechanisms, 
off-shell effects and decays of vector bosons and is fully differential in kinematics variables 
of final state leptons and jets. Contributions of massive loops are either 
treated approximately in NLO QCD, as in case 
of the $Z$-boson pair production, or completely omitted as in the case of $W$-pair production, since 
the corresponding exact computations are currently not technically feasible. 
In the case of the $Z$-pair production we construct an expansion of relevant 
amplitudes in $1/m_t$ and argue that the results of such 
an expansion can be reliably used for phenomenology provided that the four-lepton 
invariant mass is restricted to below the $2 m_t$ and hard gluons in the final state have transverse 
momenta below $150~{\rm GeV}$.  

We find that the $K$-factors for the interference in $Z$-pair production can be well described as a geometric mean of the 
$K$-factors for the background and the signal, $K_{\rm intf} \approx \sqrt{ K_{\rm bkgd} K_{\rm signal}}$.  This 
relation between the $K$-factors seems to hold both locally and globally, except in the region below and around the $2 m_Z$ threshold, where the interference $K$-factor significantly exceeds the $K$-factors for the signal and the background. This feature appears to be driven by the Higgs interference with massless prompt production amplitudes, which dominate the interference in this region.

It is interesting to 
point out that, in the Higgs signal bin $\m4l \sim m_H$, the irreducible background $gg \to ZZ$  is about one percent of the signal while the interference contributes at the level of $0.1$ percent. Since the irreducible background is flat across the signal bin, it can be constrained experimentally from side bands. At the same time, since the cross section  for Higgs boson production in gluon fusion is currently computed with a few percent precision, the interference contribution needs to receive 
a $K$-factor of more than  $10$ to become relevant. Given that the NLO interference $K$-factor stays 
close to $K \sim 2.5$ below the $2 m_Z$ threshold, 
the required enhancement is highly improbable.

For $WW$ production, we compute the QCD corrections to the interference taking into account the first two (massless) quark generations.
We find that the interference $K$-factor in that case is larger than $K$-factors for both the signal and the background.
We note, however, that this result is incomplete since for $WW$ production the contribution of top and bottom quarks
to the interference is significant.  Computation of NLO QCD corrections to the $gg \to WW$ amplitude for massive internal quarks
is an interesting challenge that we leave for future investigation.

\newpage
{\bf Acknowledgments}

We are grateful to S. Pozzorini
and  J. Lindert for helping us with checks against  {\tt OpenLoops}~\cite{Cascioli:2011va}. 
The research reported in this paper
is partially supported by BMBF grant 05H15VKCCA. F.C. and K.M. wish to thank the Kavli Institute for Theoretical Physics for hospitality
while part of this work was carried out.

{\bf Note added} When this paper was being written, Ref.~\cite{Campbell:2016ivq} appeared where 
the NLO QCD corrections to the interference in $gg \to ZZ$ were studied as well. In contrast 
to our paper,  in Ref.~\cite{Campbell:2016ivq} the production of on-shell $Z$-bosons above the 
threshold is studied  and their decays are not considered. At the same time, the attempt is made 
in Ref.~\cite{Campbell:2016ivq} to extrapolate the $1/m_t$-expansion above the $2m_t$ threshold 
using conformal mapping and Pad\'e approximants.  The results of Ref.~\cite{Campbell:2016ivq}  
suggest a close relation between $K$-factors of the signal and the interference other than in the region around the $2m_Z$ threshold, in qualitative agreement 
with what we observe in this paper.

\bibliography{ggHVVpaper}
\end{document}